\documentclass[12pt,epsf]{article}
\usepackage{graphicx,amsmath,amssymb}
\begin{document}
%%%%%%%%%%%%%%%%%%%%%%%%%%%%%%%%%%%%%%%%%%%

\def\a{\alpha}
\def\b{\beta}
\def\c{\varepsilon}
\def\d{\delta}
\def\e{\epsilon}
\def\f{\phi}
\def\g{\gamma}
\def\h{\theta}
\def\k{\kappa}
\def\l{\lambda}
\def\m{\mu}
\def\n{\nu}
\def\p{\psi}
\def\q{\partial}
\def\r{\rho}
\def\s{\sigma}
\def\t{\tau}
\def\u{\upsilon}
\def\v{\varphi}
\def\w{\omega}
\def\x{\xi}
\def\y{\eta}
\def\z{\zeta}
\def\D{\Delta}
\def\G{\Gamma}
\def\H{\Theta}
\def\L{\Lambda}
\def\F{\Phi}
\def\P{\Psi}
\def\S{\Sigma}

\def\o{\over}
\def\beq{\begin{eqnarray}}
\def\eeq{\end{eqnarray}}
\newcommand{\gsim}{ \mathop{}_{\textstyle \sim}^{\textstyle >} }
\newcommand{\lsim}{ \mathop{}_{\textstyle \sim}^{\textstyle <} }
\newcommand{\vev}[1]{ \left\langle {#1} \right\rangle }
\newcommand{\bra}[1]{ \langle {#1} | }
\newcommand{\ket}[1]{ | {#1} \rangle }
\newcommand{\EV}{ {\rm eV} }
\newcommand{\KEV}{ {\rm keV} }
\newcommand{\MEV}{ {\rm MeV} }
\newcommand{\GEV}{ {\rm GeV} }
\newcommand{\TEV}{ {\rm TeV} }
\def\diag{\mathop{\rm diag}\nolimits}
\def\Spin{\mathop{\rm Spin}}
\def\SO{\mathop{\rm SO}}
\def\O{\mathop{\rm O}}
\def\SU{\mathop{\rm SU}}
\def\U{\mathop{\rm U}}
\def\Sp{\mathop{\rm Sp}}
\def\SL{\mathop{\rm SL}}
\def\tr{\mathop{\rm tr}}

\def\IJMP{Int.~J.~Mod.~Phys. }
\def\MPL{Mod.~Phys.~Lett. }
\def\NP{Nucl.~Phys. }
\def\PL{Phys.~Lett. }
\def\PR{Phys.~Rev. }
\def\PRL{Phys.~Rev.~Lett. }
\def\PTP{Prog.~Theor.~Phys. }
\def\ZP{Z.~Phys. }

%%%%%%% added by Fumi %%%%%%%%%%
% FROM HERE
%\newcommand{\beq}{\begin{equation}}   
%\newcommand{\eeq}{\end{equation}}
\newcommand{\bea}{\begin{eqnarray}}   
\newcommand{\eea}{\end{eqnarray}}
\newcommand{\bear}{\begin{array}}  
\newcommand {\eear}{\end{array}}
\newcommand{\bef}{\begin{figure}}  
\newcommand {\eef}{\end{figure}}
\newcommand{\bec}{\begin{center}}  
\newcommand {\eec}{\end{center}}
\newcommand{\non}{\nonumber}  
\newcommand {\eqn}[1]{\beq {#1}\eeq}
\newcommand{\la}{\left\langle}  
\newcommand{\ra}{\right\rangle}
\newcommand{\ds}{\displaystyle}
\def\SEC#1{Sec.~\ref{#1}}
\def\FIG#1{Fig.~\ref{#1}}
\def\EQ#1{Eq.~(\ref{#1})}
\def\EQS#1{Eqs.~(\ref{#1})}
\def\TEV#1{10^{#1}{\rm\,TeV}}
\def\GEV#1{10^{#1}{\rm\,GeV}}
\def\MEV#1{10^{#1}{\rm\,MeV}}
\def\KEV#1{10^{#1}{\rm\,keV}}
\def\lrf#1#2{ \left(\frac{#1}{#2}\right)}
\def\lrfp#1#2#3{ \left(\frac{#1}{#2} \right)^{#3}}
\def\REF#1{Ref.~\cite{#1}}
\newcommand{\osc}{{\rm osc}}
\newcommand{\ed}{{\rm end}}
\def\dda#1{\frac{\partial}{\partial a_{#1}}}
\def\ddat#1{\frac{\partial^2}{\partial a_{#1}^2}}
\def\dd#1#2{\frac{\partial #1}{\partial #2}}
\def\ddt#1#2{\frac{\partial^2 #1}{\partial #2^2}}
\def\lrp#1#2{\left( #1 \right)^{#2}}
% UNTIL HERE

%%%%%%%%%%%%%%%%%%%%%%%%%%%%%%%%%%%%%%%%%%%%%%%%%%%%%%%%%%%%%%%%%%%

\baselineskip 0.7cm

\begin{titlepage}

\begin{flushright}
UT-14-09\\
TU-957\\
IPMU14-0054\\
\end{flushright}

\vskip 1.35cm
\begin{center}
{\large \bf 
The 3.5 keV X-ray line signal from decaying moduli \\
with low cutoff scale
}
\vskip 1.2cm
Kazunori Nakayama$^{a,c}$,
Fuminobu Takahashi$^{b,c}$
and 
Tsutomu T. Yanagida$^{c}$

\vskip 0.4cm

{\it $^a$Department of Physics, University of Tokyo, Tokyo 113-0033, Japan}\\
{\it $^b$Department of Physics, Tohoku University, Sendai 980-8578, Japan}\\
{\it $^c$Kavli Institute for the Physics and Mathematics of the Universe (WPI), TODIAS, University of Tokyo, Kashiwa 277-8583, Japan}

\vskip 1.5cm

\abstract{
The recent unidentified 3.5 keV X-ray line signal can be explained by decaying moduli dark matter
with a cutoff scale one order of magnitude smaller than the Planck scale. We show that
such modulus field with the low cutoff scale follows a time-dependent potential minimum and its
abundance is reduced by the adiabatic suppression mechanism. As a result the modulus abundance can naturally be
consistent with the observed dark matter abundance without any fine-tuning of the initial oscillation
amplitude.
}
\end{center}
\end{titlepage}

\setcounter{page}{2}

%%%%%%%%%%%%%%%%%%%%%%%%%%%
%\section{Introduction}
%%%%%%%%%%%%%%%%%%%%%%%%%%%

Recently, two groups reported strong evidence for the unidentified X-ray line at about $3.5$ keV from 
various galaxy clusters and the Andromeda galaxy~\cite{Bulbul:2014sua,Boyarsky:2014jta}.
While there may be systematic uncertainties relevant to these observations,
the reported X-ray line could be due to the decay of 7 keV dark matter. It is worth noting that
this interesting possibility was pointed out many years ago by Kawasaki and one of the present authors (TTY)
in Ref.~\cite{Kawasaki:1997ah} (see also Refs.~\cite{Hashiba:1997rp,Kusenko:2012ch}), 
where they studied decay of a light moduli field into photons in the gauge mediated supersymmetry (SUSY) breaking.
After the discovery of the $3.5$\,keV X-ray line, there appeared various possibilities of producing
the X-ray line by dark matter (DM)~\cite{Higaki:2014zua,Ishida:2014dlp,Jaeckel:2014qea,Lee:2014xua,Finkbeiner:2014sja,Krall:2014dba,Kong:2014gea,Frandsen:2014lfa}.
In fact, the DM with mass and lifetime 
\begin{gather}
	m_{\phi} \simeq 7\,{\rm keV}, \\
	\tau_{\phi} \simeq 2\times 10^{27} - 2\times 10^{18}\,{\rm sec},
\end{gather}
can explain the excess where $\phi$ denotes the DM particle~\cite{Boyarsky:2014jta}. Note that, if the DM decays
into a pair of photons, a factor $2$ needs to be multiplied with the lifetime. 
The task for the model building is how to realize such small DM mass, long lifetime as well as the correct DM abundance.

Let us focus on light moduli dark matter coupled to photons
with a relatively low cutoff scale
 $M$~\cite{Kawasaki:1997ah,Hashiba:1997rp,Kusenko:2012ch,Higaki:2014zua,Jaeckel:2014qea,Lee:2014xua,Krall:2014dba}:
\begin{equation}
	\mathcal L = \frac{\phi}{4M}F_{\mu\nu}F^{\mu\nu}~~~{\rm or}~~~ \frac{\phi}{4M}F_{\mu\nu}{\tilde F}^{\mu\nu},
\end{equation}
where $\phi$ is a real scalar for the former, whereas it is a real pseudo-scalar for the latter.
Since the moduli generally obtain mass from SUSY breaking effect, the moduli with 
mass of $\mathcal O$(keV) naturally appears in the gauge-mediated SUSY 
breaking models, and its cosmological problem was studied in 
Refs.~\cite{Kawasaki:1997ah,Hashiba:1997rp,Kasuya:2001tp}.
Assuming that the modulus decays mainly into photons through the above operator,
the modulus lifetime is given by
\begin{equation}
	\tau_\phi = \left( \frac{m_\phi^3}{64\pi M^2} \right)^{-1}
	\simeq 4\times 10^{27}\,{\rm sec}\left( \frac{7\,{\rm keV}}{m_\phi} \right)^3\left( \frac{M}{10^{17}\,{\rm GeV}} \right)^2.
\end{equation}
Thus, the reported 3.5 keV X-ray line  can be explained by the decaying moduli DM with $m_\phi \simeq 7$ keV, if the cutoff scale $M$ is about one order of magnitude smaller than the Planck scale, $M_P \simeq 2.4\times 10^{18}\,$GeV.\footnote{ 
The suggested value of the cut-off scale is within the range naturally expected in 
string theory~\cite{Higaki:2014zua}.
}

The potential problem of the light moduli dark matter scenario is that the modulus abundance may exceed the
observed DM abundance by many orders of magnitude, which is known as the
 cosmological moduli problem~\cite{Coughlan:1983ci,Banks:1993en,Kawasaki:1997ah,Kasuya:2001tp}.
  In this short note we point out that the moduli field
 with the low cut-off scale follows a time-dependent minimum and its subsequent oscillation amplitude is significantly 
reduced by the so called adiabatic suppression mechanism~\cite{Linde:1996cx,Takahashi:2010uw,Nakayama:2011wqa,Nakayama:2011zy,Nakayama:2012mf},  and 
it can actually account for DM for appropriate choice of inflation energy scale.

The point is that the modulus field obtains a large Hubble mass of $\mathcal O(10) \times H$ with $H$ being  the Hubble scale,  through the  following coupling with the inflaton,\footnote{
	Note that, if $\phi$ respects a shift symmetry, $\phi \to \phi + i \alpha$ with $\alpha$ being a real parameter, the adiabatic
	suppression does not apply to the imaginary component of $\phi$.
}
\begin{equation}
	K \supset \frac{b^2}{M^2} |\phi|^2 | I |^2,
\end{equation}
where $I$ denotes the inflaton superfield and $b$ is an order one numerical coefficient.
Here we assume that the modulus field universally couples to other fields with the cutoff scale $M \sim 0.1 M_P$.
In such a case, the amplitude of coherent oscillations of the modulus field is much suppressed because
of the  adiabatic suppression 
mechanism.

The coherent oscillation of the moduli in this case is induced only just after inflation when the adiabaticity of the modulus dynamics is temporarily violated, and as a result, its abundance is given by~\cite{Nakayama:2011wqa,Nakayama:2011zy,Nakayama:2012mf}\footnote{
The moduli abundance is {\it not} exponentially suppressed if one follows the modulus dynamics
from during inflation. This is because the end of inflation necessarily breaks the adiabaticity, which induces the modulus oscillations~\cite{Nakayama:2011wqa}.
}
\begin{equation}
\begin{split}
	\frac{\rho_\phi}{s} & \sim \frac{1}{8}T_{\rm R}\left( \frac{\phi_0}{M} \right)^2 \left( \frac{m_\phi}{cH_{\rm inf}} \right) \\
	&\sim 4\times 10^{-10}\,{\rm GeV}\left( \frac{T_{\rm R}}{10^4\,{\rm GeV}} \right)
	\left( \frac{\phi_{0}}{M} \right)^2
	\left( \frac{M}{10^{17}\,{\rm GeV}} \right)^2
	 \left( \frac{m_\phi}{7\,{\rm keV}}\right)
	 \left( \frac{10^7\,{\rm GeV}}{H_{\rm inf}}\right),
	 \label{abund}
\end{split}
\end{equation}
where $T_{\rm R}$ is the reheating temperature after inflation, $H_{\rm inf}$ is the Hubble scale during inflation
and $c \sim M_P/M$.
Note that we need $T_{\rm R} \lesssim \sqrt{m_\phi M_P} \sim 10^6$\,GeV in order for the adiabatic suppression to work.
Otherwise, the universe is radiation dominated at the onset of moduli oscillation,
during which the Hubble mass of the moduli is relatively suppressed~\cite{Kawasaki:2011zi}.\footnote{
Even smaller cut-off scale for the interactions between the modulus and the standard model particles is needed
for the adiabatic suppression mechanism to work in this case.
}
Thus the light moduli with a keV mass  can be a dominant DM for appropriate choices of $T_{\rm R}$ and $H_{\rm inf}$,
without severe fine-tuning of the initial oscillation amplitude, contrary to the conventional wisdom.  

Several comments are in order.
Low scale gauge mediation models are consistent with the recent LHC data with 126\,GeV Higgs boson,
and may be tested at the 14\,TeV LHC~\cite{Hamaguchi:2014sea}.
The gravitino thermal production can be suppressed if sparticles are as heavy as the reheating temperature.
Another option is to introduce the late time entropy production:
the dilution of the gravitino by a factor of $\mathcal O(100)$ is sufficient to avoid the gravitino overproduction~\cite{Hamaguchi:2014sea}.
In this case, the moduli abundance (\ref{abund}) is also reduced by the same dilution factor,
but it can be compensated by the increase of the reheating temperature.
Nonthermal leptogenesis may work if one assumes a mild degeneracy among right-handed neutrino masses.
Finally, low inflation scale of $H_{\rm inf} \sim 10^7$\,GeV is consistent with the SUSY new inflation scenario~\cite{Nakayama:2012dw}.

%%%%%%%%%%%%%%%%%%%%%%%%%%%%%%%%%%%%%%%%%%%%
\section*{Acknowledgments}
%%%%%%%%%%%%%%%%%%%%%%%%%%%%%%%%%%%%%%%%%%%%
This work was supported by the Grant-in-Aid for Scientific Research on
Innovative Areas (No. 21111006  [KN and FT],  No.23104008 [FT], No.24111702 [FT]),
Scientific Research (A) (No. 22244030 [KN and FT], 21244033 [FT], 22244021 [TTY]),  JSPS Grant-in-Aid for
Young Scientists (B) (No.24740135) [FT], and Inoue Foundation for Science [FT].  This work was also
supported by World Premier International Center Initiative (WPI Program), MEXT, Japan.

%%%%%%%%%%%%%%%%%%%%%%%%%%%%%%%%%%%%%%%%%%%%

\end{document}